%% file: nonmonotonic.tex
\documentclass[aps,prb,twocolumn,superscriptaddress,showpacs,citeautoscript]{revtex4-1}
\usepackage{graphicx}
\usepackage{amsmath,mathrsfs,amsfonts}
\usepackage{color}
\usepackage{placeins}
\usepackage{epstopdf}
\usepackage{subfigure}
\usepackage{hyperref}
\usepackage{grffile}
\usepackage{enumerate}

\begin{document}
\newcommand{\<}{\langle}
\renewcommand{\>}{\rangle}
\newcommand{\Lmin}{L_{\rm min}}
\newcommand{\vv}[1]{\langle #1 \rangle}
\newcommand{\beq}{\begin{equation}}
\newcommand{\eeq}{\end{equation}}
\newcommand{\ra}{\rightarrow}
\newcommand{\Ra}{\Rightarrow}
\newcommand{\Lra}{\Longrightarrow}
\newcommand{\Sra}{\Shortrightarrow}
\newcommand{\abs}[1]{ \left| #1 \right| }
\newcommand{\pa}[1]{ \left( #1 \right) }
\newcommand{\mas}[1]{ \left\{ #1 \right\} }
\newcommand{\hak}[1]{ \left[ #1 \right] }
\newcommand{\txt}[1]{\textrm{#1}}
\newcommand{\bmat}{\begin{displaymath}}
\newcommand{\emat}{\end{displaymath}}
\newcommand{\nb}{\hat{n}_{\mathbf{r}}}
\newcommand{\vrand}{v_{\mathbf{r}}}
\newcommand{\rup}{\mathbf{r}}
\newcommand{\phihat}{\hat{\Phi}}
\newcommand{\tehat}{\hat{\Theta}}
\newcommand{\Jtau}{J^\tau_{\rup}}
\newcommand{\ampfluc}{\delta{n_{\rup}}}
\newcommand{\ampflucop}{\delta{\hat{n}_{\rup}}}
\newcommand{\rhos}{\rho_{\text{s}}}
\newcommand{\tr}[1]{\text{Tr}\mas{#1}}
\newcommand{\kc}{K_{\text{c}}}
\newcommand{\deriveat}[2]{ \left. #1 \right|_{#2} }
\newcommand{\todo}[1]{\textbf{\textsc{\textcolor{red}{(TODO: #1)}}}}
\newcommand{\correct}[1]{\noindent\textbf{\textsc{\textcolor{red}{#1}}}}
\newcommand{\vect}[1]{\mathbf{#1}}
\newcommand{\rb}{\mathbf{r}}
\newcommand{\kb}{\mathbf{k}}
\newcommand{\tc}{T_{\text{c}}}
\newcommand{\tcone}{T_{\text{c},1}}
\newcommand{\tctwo}{T_{\text{c},2}}
\newcommand{\ec}{E_{\text{c}}}
\newcommand{\cvv}{c_V}
\newcommand{\cdd}{c_U}
\newcommand{\lvv}{\lambda}
\newcommand{\ldd}{\xi_L}

\title{Fluctuation-induced first-order phase transitions in type-1.5
  superconductors in zero external field} 
\date{\today} 
\author{Hannes Meier} \affiliation{Department of Theoretical Physics,
  KTH Royal Institute of Technology, SE-106 91 Stockholm, Sweden}
\author{Egor Babaev} \affiliation{Department of Theoretical Physics,
  KTH Royal Institute of Technology, SE-106 91 Stockholm, Sweden}
\author{Mats Wallin} \affiliation{Department of Theoretical Physics,
  KTH Royal Institute of Technology, SE-106 91 Stockholm, Sweden}
\begin{abstract}
  In a single-component Ginzburg-Landau model which possesses
  thermodynamically stable vortex excitations, the zero-field
  superconducting phase transition is second order even when
  fluctuations are included.  Beyond the mean-field approximation the
  transition is described in terms of proliferation of vortex loops.
  Here we determine the order of the superconducting transition in an
  effective 3D vortex-loop model for the recently proposed multiband
  type-1.5 superconductors.  The vortex interaction is nonmonotonic,
  i.e., exponentially screened and attractive at large separations,
  and short-range repulsive.  We show that the details of the vortex
  interaction, despite its short-range nature, can lead to very
  different properties of the superconducting transition than found in
  type-1 and type-2 systems. Namely, the type-1.5 regime with
  nonmonotonic intervortex interaction can have a first-order
  vortex-driven phase transition not found in the single-band case.
\end{abstract}
\maketitle

\section{Introduction}

The order of the zero-field superconducting transition has been
studied in many works in the usual single-component type-1 and type-2
superconductors.  Halperin, Lubensky, and Ma established that in
extreme type-1 superconductors the gauge field fluctuations render the
superconducting phase transition first order
\cite{HalperinLubenskyMa,COLEMANWEINBERG}.  In the opposite limit of
extreme type-2 systems, Dasgupta and Halperin \cite{DasguptaHalperin}
demonstrated that the superconducting transition is continuous and in
the universality class of the inverted-3DXY model.  The different
nature of the superconducting phase transition in this limit is
revealed by a duality mapping
\cite{DasguptaHalperin,Peskin1978122,Thomas1978513}, which
demonstrates that the phase transition is driven by proliferation of
vortex-loop fluctuations.

While the extreme type-1 and type-2 limiting cases are well
investigated, the value of Ginzburg-Landau parameter
$\kappa=\lambda/\xi$ at which the phase transition changes from second
to first order is much harder to establish. The attempted analytical
approaches \cite{KleinertKappa} are based on approximations that are
unfortunately not controllable, in contrast to the well controllable
duality mapping in the London limit
\cite{DasguptaHalperin,Peskin1978122,Thomas1978513}.  The most
reliable information to date comes from numerical simulations. The
largest Monte Carlo simulations performed so far
\cite{SudboKappa,SudboKappa2} claim that the tricritical
$\kappa_{\text{tri}}=\pa{0.76\pm0.04}/\sqrt2$ is slightly smaller than
the critical $\kappa_c=1/\sqrt2$, which separates the type-1 regime
with thermodynamically unstable vortices and the type-2 regime with
thermodynamically stable vortices.  In these works it is claimed that
even in the weakly type-1 regime where the vortex interaction is
purely attractive and vortices are not thermodynamically stable, the
phase transition can be continuous.\footnote{The interaction is purely
  attractive between type-1 vortices only in the continuum limit
  \cite{SudboKappa,SudboKappa2}; for lattice Ginzburg-Landau model
  there is always contact repulsion between vortex lines. The contact
  repulsion can, for entropic reasons make the effective interaction
  purely repulsive \cite{coppersmith,zaanen}. In
  Refs.~\cite{SudboKappa,SudboKappa2} it was discussed that the effect
  can result in a continuous phase transition for vortices with bare
  attractive interaction and contact repulsion. The difference in our
  case is that we consider the regime with thermodynamically stable
  vortices with repulsion, which is not contact, and no other degrees
  of freedom present in the model.}

Recently it has been proposed that in multicomponent superconductors
there is a new regime that falls outside the type-1/type-2
classification.  Such materials are described by theories with
multiple superconducting components, e.g., by Ginzburg-Landau theory
of the form
\begin{equation}
  F=\sum_{a=1,2}\frac{1}{2} |(i\nabla -e{\bf A})\psi_a|^2 + V(|\psi_a|,\theta_1-\theta_2) 
  +\frac{(\nabla\times {\bf A})^2}{2}
\end{equation}
where $\psi_a=|\psi_a|e^{i\theta_a}$ are superconducting components,
$V$ is a collection of potential terms, and ${\bf A}$ is the vector
potential.  Such systems have multiple coherence lengths $\xi_a$.  For
detailed discussion of the definitions of coherence lengths in the
presence of inter-component coupling, see
Ref.~\onlinecite{JohanGeneralReview}.  In type-1.5 regimes some of the
coherence lengths are larger and some are smaller than the magnetic
field penetration length $\lambda$
\cite{Babaev.Speight:05,JohanProximityEffect,JohanGeneralReview}.  The
different coherence lengths can originate from the existence of
different superconducting gaps in different bands \cite{silaevmicro},
or superconducting states breaking multiple symmetries
\cite{2011PhRvB..84m4518C,2014PhRvB..90f4509A}.

In what follows we focus on the two-band case. It has been shown that
thermodynamically stable double-core vortices exist in the regime
where $\xi_1<\sqrt2\lambda<\xi_2$.  In 2D such vortices asymptotically
have an interaction of the form
\cite{Babaev.Speight:05,JohanProximityEffect,JohanGeneralReview}
\beq V(r) \sim m^2 K_0\pa{r/\lambda} - q_1^2 K_0\pa{r/\xi_1} - q_2^2
K_0\pa{r/\xi_2}
\label{Eq:2Dinter}
\eeq  
where $m,q_1,q_2$ are system dependent coefficients and $K_0$ is a
modified Bessel function.  The first term in Eq.~\eqref{Eq:2Dinter}n
with range $\lambda$ originating from the magnetic and current-current
interaction is repulsive for two vortices with like vorticity and
attractive otherwise. The second and third terms are attractive with
range $\xi_L=\max\mas{\xi_1,\xi_2}$, and originate from core-core
interaction for vortices with two cocentered overlapping cores in the
two superconducting components. Consequently in type-1.5 regime with
$\xi_L>\lambda$ the interaction is short-range repulsive due to the
first term, while at the longer range it is exponentially screened and
attractive due to the core-core attraction. We will refer to the
core-core attraction as intermediate-range attractive to emphasize
that all interactions here are exponentially screened.  In contrast to
the type-1 regime, type-1.5 systems have thermodynamically stable
vortex excitations, while in contrast to the type-2 regime the
intermediate-range intervortex forces are attractive.  Therefore the
nature of the superconducting phase transition in the type-1.5 systems
cannot be deduced from known cases of single-component
superconductors.  Currently the problems of type-1.5 superconductivity
is a subject of intense experimental research on materials where
vortex clusters were observed
\cite{MoshchalkovType15,ray,
  GutierrezHallProbe,NishioSQUID,PhysRevB.81.214501}.

Similar to the single-component case, in two-band systems it is
difficult to advance analytically in a controllable way away from
extremely type-2 regimes, in particular using duality arguments.
Nonetheless one can identify a limit in the type-1.5 regime where
certain simplifying assumptions can be made.  That is, consider a
two-band superconductor with relatively strong interband coupling made
of a strongly type-2 component and a type-1 component with a much
lower ground state density.  This condition implies that vortex
excitations are expected to drive the phase transition. Yet in
contrast to type-2 superconductors, the vortices will feature a small
attractive tail in the interaction.  In a regime with relatively
strong interband coupling, vortices can be approximated as objects
with no fluctuating internal structure, and, under certain conditions,
multibody forces between type-1.5 vortices can be neglected
\cite{JohanClustersMultibodyForces}.  Then the composite vortices can
be seen as charged point particles interacting via a sum of screened
Coulomb potential terms.

In this paper we study the 3D vortex-loop driven finite-temperature
superconducting phase transition in zero field for a model of a
type-1.5 superconductor.  The main task is to investigate the order of
the superconducting transition.  We propose an effective model for
composite vortices with a nonmonotonic length scale dependence of the
vortex interaction.  We study this model by classical
finite-temperature Monte Carlo (MC) simulation and finite-size scaling
methods in order to classify the order of the transition.  From the
U(1) symmetry of the superconducting order parameter an inverted 3DXY
transition is the expected result for a system with thermodynamically
stable vortices, but instead we obtain first-order transitions in the
cases involving a nonmonotonic vortex interaction that we tested.
This result differs qualitatively from the single-band systems where
the zero field transition driven by thermodynamically stable vortices
is considered to be always continuous.

\section{Generalized effective vortex loop model}

For the type-1.5 regime in the limit outlined above, fluctuations near
the phase transition can be expected to be described by a generalized
3D vortex loop model that we will now formulate.  In a 3D system
vortex lines form closed loops, and on a lattice the vortex degrees of
freedom become directed integer link current variables $q_i^\sigma$,
where $\sigma=\hat{x},\hat{y},\hat{z}$ are the unit lattice vectors
connecting the site $i$ with its neighbor $i+\sigma$ on a simple cubic
lattice with vertices $i=1,\dots,L^d$.  The numerical lattice constant
is set to unity.  The functions $K_0(r/\lambda)$ in
Eq.~\eqref{Eq:2Dinter} generalize to 3D Yukawa interactions,
represented on a lattice with periodic boundary conditions by lattice
Green's functions
\beq 
Y_{ij}=Y\pa{\frac{\abs{\vect{r}_i-\vect{r}_j}}{\lambda_Y}}
=\frac{c_Y}{L^d}\sum_{\vect{k}}
\frac{\cos\pa{\vect{k}\cdot\pa{\vect{r}_i-\vect{r}_j}}}{6-
  \sum_\sigma{2\cos\pa{k_\sigma}}+ \lambda_Y^{-2}}
\label{Eq:VVInteract} 
\eeq 
where $c_Y$ is a real coupling constant. The 3D counterpart of
Eq.~\eqref{Eq:2Dinter} is then given by the vortex line Hamiltonian
\beq 
H = \sum_{i,j,\sigma} \frac{1}{2} q_i^\sigma V_{ij} q_j^\sigma +
\sum_{i,j,\sigma} \frac{1}{2} \abs{q_i^\sigma} U_{ij} \abs{q_j^\sigma}
\label{Eq:ModelHam}
\eeq  
where both $U_{ij}$ and $V_{ij}$ shall have the form of
Eq.~\eqref{Eq:VVInteract}.  The first term corresponds to
$m^2K_0\pa{r/\lambda}$ in Eq.~\eqref{Eq:2Dinter} with
$\lambda_V=\lambda$ in Eq.~\eqref{Eq:VVInteract}. Thus $V_{ij}$
mediates the screened Coulomb interaction of the composite vortex
lines as obtained for a two-component 3D superconductor with range set
by the London penetration depth $\lambda$.  The second term
corresponds to the slowest decaying density interaction
$-q_L^2K_0\pa{r/\xi_L}$ in Eq.~\eqref{Eq:2Dinter}, which is always
attractive ($\cdd<0$) and of exactly the same form as $V_{ij}$ with
range $\lambda_U=\xi_L$.  The faster decaying component has been
ignored meaning that its range and amplitude are assumed to be
sufficiently small.  This model is highly simplified and neglects
amplitude fluctuations, additional core-energy contributions, and
core-core interactions between perpendicular line segments.  While
such effects can in principle be included in the model to reach
accurate description of a given material, we here focus on properties
of the effective model and leave more detailed investigations for
future work.

For a weak attractive part $\abs{c_U} \ll c_V$ and $\xi_L\leq
\lambda$, Eq.~\eqref{Eq:ModelHam} is similar to a type-2
superconductor, and from the U(1) symmetry of the model it is expected
that the transition from the ordered low-temperature phase to the
disordered high-temperature phase is a second-order phase transition
belonging to the inverted 3DXY universality class.  However for a
general choice of parameters in Eq.~\eqref{Eq:ModelHam} such an 
{\em a priori} assertion is not possible.  The order of the transition must in
general be determined by simulations or by other means.

Next we discuss the choice of parameter values.  The possible
parameter choices are restricted by a stability criterion in order to
represent a valid description for multicomponent superconductors.
That is, the coefficients $\cdd<0<\cvv$ and the ranges $\lvv,\ldd$ in
Eq.~\eqref{Eq:ModelHam} must be chosen such that the lowest energy
state is the vortex free state with all the $q_i^\sigma = 0$.  At the
parameters we will consider the minimum of the vortex interaction
comes from the attraction energy between nearest neighbor link
variables with opposite sign.  A candidate low energy state is thus
given by a N\'eel-type stacked loop configuration on a cubic lattice
such as
$q^z\pa{\rb}=\pa{-1}^{x+y+z},q^y\pa{\rb}=-q^z\pa{\rb},q^x\pa{\rb}=0$.
The energy of the stacked state can be calculated from the Hamiltonian
in Fourier space, $H = \frac{1}{2L^d} \sum\limits_{\sigma,{\bf k}}
\hak{\tilde{V} ({\bf k}) | \tilde{q}^\sigma({\bf k}) |^2 +
  \tilde{U}({\bf k}) | \tilde{Q}^\sigma({\bf k}) |^2 }$, where
$Q^\sigma(\rb)=|q^\sigma(\rb)|$, which gives $E = L^d [
\tilde{V}(\pi,\pi,\pi) + \tilde{U}(0,0,0) ]= L^d [ \cvv/(12+\lvv^{-2})
+\cdd\ldd^2 ]$.  The boundary of stable parameters is identified by
setting this energy to zero to make the stacked state degenerate with
the vortex-free vacuum state.  The parameters used in the model must
thus satisfy
\beq
\frac{\cvv}{12+\lvv^{-2}}  +\cdd\ldd^2 > 0
\label{StabReqStable}
\eeq
To investigate the different types of behavior of the model in
Eq.~\eqref{Eq:ModelHam} we focus on several different parameter
regimes:

(1) Screened repulsive parameters (SR):
$\ldd=\lvv=0.5,\cvv=41,\cdd=-2.5$.  The attractive coefficient $\cdd$
is small compared to $\cvv$ yielding a net repulsive interaction
between vortex segments with equal vorticity, thus representing a
two-band type-2 superconductor.  The transition in this model is
therefore expected to belong to the inverted-3DXY universality class,
which will be verified below.

(2) Nonmonotonic parameters (NM): $\ldd=1, \lvv=0.5,
\cvv=41,\cdd=-2.5$.  The range $\ldd$ of the attractive part has been
increased compared to the SR case yielding an effectively
nonmonotonic interaction with a net repulsion at short length scales
and a net attraction at intermediate length scales between equal
vorticities.  For these values of $\ldd, \lvv, \cdd$, the choice
$\cvv=41$ is within the stability requirement $\cvv>40$ given by
Eq.~\eqref{StabReqStable}.  This regime gives a simplified effective
model for vortex loops in type-1.5 superconductors.

(3-4) Screened repulsive parameters with enhanced attraction (SR10,
SR1024999): $\lvv=\ldd=0.5,\cvv=41$ and $\cdd=-10,-10.24999$,
respectively, which are close to the minimum allowed value $-10.25$
set by Eq.~\eqref{StabReqStable}.  For the SR1024999 parameters the
vortex free state and the staggered configuration are almost
degenerate in energy.  In ordinary superconductors the energy of such
vortex configurations contain contributions from nonpairwise forces,
which are not present in our model.  In this parameter regime the
model is not representative for currently known ordinary
superconductors and has mainly theoretical interest.

\section{Calculated quantities and scaling arguments}

Destruction of superconductivity in systems with thermodynamically
stable vortices is associated with proliferation of vortex loop
fluctuations.  For finite $\lambda$ we can assume an ensemble where
number fluctuations of the vortex lines are included at finite energy
cost. The proliferation of vortex loops at the phase transition is
signaled by fluctuations in the winding numbers
\beq 
W_\sigma = \frac{1}{L} \sum_i q_i^\sigma
\label{Eq:W}
\eeq 
The singular behavior at a second-order phase transition is described
by the finite size scaling ansatz
\beq
\<W_\sigma^2\> = \tilde{W}^2(L^{1/\nu}t)
\label{Eq:Wfluctuation}
\eeq
where $\tilde{W}^2$ is a scaling function, $L$ is the system size,
$t=T/T_c-1$, and $\nu$ is the correlation length critical exponent.
This means that curves of MC simulation data of $\<W_\sigma^2\>$ vs
temperature $T$ for different system sizes $L$ will intersect at
$T=T_c$.  The derivative scales as 
$\partial \<W_\sigma^2\>/\partial T \sim L^{1/\nu}$
at the transition.  For the 3DXY universality class the critical
exponent for the correlation length is $\nu\approx 0.671$ and for the
heat capacity $\alpha=2-d\nu\approx-0.015$ \cite{Hasenbusch}.

In the vicinity of a first-order transition the two different phases
coexist, the correlation length is finite, and scaling given by
Eq.~\eqref{Eq:Wfluctuation} is not fulfilled.  Precisely at $\tc$ the
system is equally probable to be in either of the phases.  In
simulations the internal energy histogram $P(e)$, where $e=H/L^d$ is
the energy density, shows a double-peak structure centered around the
two characteristic internal energy values $H_1,H_2$.  The free-energy
barrier given by $\Delta F_L =
(1/\beta)\log\hak{{P_{\text{max}}}/{P_{\text{min}}}}$ increases with
system size and behaves asymptotically as $L^{d-1}$
\cite{LeeKosterlitz}.  For a second-order transition the double-peak
structure disappears in the thermodynamic limit.

The presence of a double-peak structure in the energy histogram is not
enough to distinguish between a first- and a second-order transition.
For a first-order transition it is also required that the latent heat
$\Delta H=H_1-H_2$ does not vanish in the thermodynamic limit.  The
latent heat contributes to the heat capacity
\beq
c_L\pa{T}=\frac{\vv{\pa{H-\vv{H}}^2}}{T^2L^d} 
\eeq 
A double-peak structure in $P(e)$ leads to a heat capacity maximum at
the transition with a leading size dependence given by $c_L^* \sim L^d
(\Delta e)^2 \sim L^d$, corresponding to a delta-function singularity
at $T_c$ for $L\to\infty$.  This is equivalent to an energy peak
separation given by $\Delta e \sim (c_L^* / L^d)^{1/2} >0$.  On the
contrary, if the transition is continuous the scaling form $c_L(T)
\approx a(t,L)t^{-\alpha}+b(t) \approx L^{\alpha/\nu}
\tilde{a}\pa{L^{1/\nu} t}+ b(t)$ holds.  This implies that the maximum
$c_L^*$ grows slower upon increasing the system size than in the case of a
first-order transition as long as $d\nu >1$, and a histogram with a
single energy peak.  In the data analysis below we sometimes find it
useful to plot the rescaled heat capacity $c_L/L^d$ which for
increasing system size should approach a constant maximum for a first
order transition and a decreasing maximum for a 3DXY transition.  To
reduce the influence of the analytic term it can be beneficial to
consider the third moment \cite{PhysRevB.67.205104}
\beq
M_{3}=\vv{H-\vv{H}}^{3}\sim\frac{\partial}{\partial T} \pa{T^{2}L^{d}c_{L}}
\eeq
This quantity exhibits two extrema around $\tc$ whose difference
$\Delta M_{3}$ scales as $\sim L^{\frac{1+\alpha}{\nu}}$ for
continuous and $\sim L^{2d}$ for discontinuous transitions. The size
dependence of $c_L^*$ and $\Delta M_{3}$ at a first-order transition
corresponds to effective exponents $\alpha=1,\nu=1/d$.

In addition a method by Challa, Landau, and
Binder\cite{PhysRevB.34.1841} which does not rely on a precision
determination of the energy histogram can be used to determine the
order of a transition. The reduced fourth order energy cumulant for
system size $L$ is
\beq 
V_L=1-\frac{\vv{H^{4}}}{3\vv{H^2}^2} 
\eeq 
For both discontinuous and continuous transitions this quantity
approaches the trivial limit $V^{*}_\infty=2/3$ for $T\neq\tc$.  For
finite-size systems a minimum $V_L<V^{*}_\infty$ is obtained at the
transition.  For second-order transitions this minimum converges
towards $V^{*}_{\infty}$ for $L\to\infty$, while for first-order
transitions the minimum approaches a nontrivial value
$V_\infty<V^{*}_\infty$ with a correction term $\sim L^{-d}$.

\section{Monte Carlo methods}

Our MC simulations use a hybrid scheme combining worm and exchange
methods that performs well both at first- and second-order phase
transitions.  The classical worm algorithm \cite{WormAlgClass}
constructs closed vortex loop fluctuations in terms of closed random
walk trajectories, and gives efficient simulation performance at a
second-order phase transition by minimizing critical slowing down of
the dynamics. In addition the replica exchange algorithm
\cite{RepExMC} is used in order to reduce the autocorrelation time and
the risk of getting stuck in metastable states, which reduces
hysteresis at first-order transitions. In our simulations we use 8-80
parallel threads. Prior to each production run the system was
equilibrated for $2^{15}$ sweeps for the SR parameters, for at least
$2^{17}$ sweeps for the NM and SR1024999 parameter sets, and for
$2^{16}$ sweeps for the SR10 set.  A MC sweep is taken to be $3L^d$
link variable updating attempts.  Equally many sweeps were done to
compute averages.  The MC trajectories were then further analyzed
using the multihistogram reweighting technique
\cite{MultiHistRew}. Error bars are obtained via reweighting of
different bootstrap realizations of the same MC trajectory.  As a
consistency check we also performed parallelized Wang-Landau
simulations \cite{WangLandau,ParallelWL} for the NM case in a finite
energy window determined by the energy expectation value of the MC
simulations.  In the WL simulations the MC moves used are two types of
closed loops, formed as attempts to insert closed elementary plaquette
loops or as straight lines that close on themselves by the periodic
boundary conditions.  Consistent results between the different
methods were obtained.

\section{Results}

\begin{figure}[tbp]
\begin{center}
\includegraphics[width=\columnwidth]{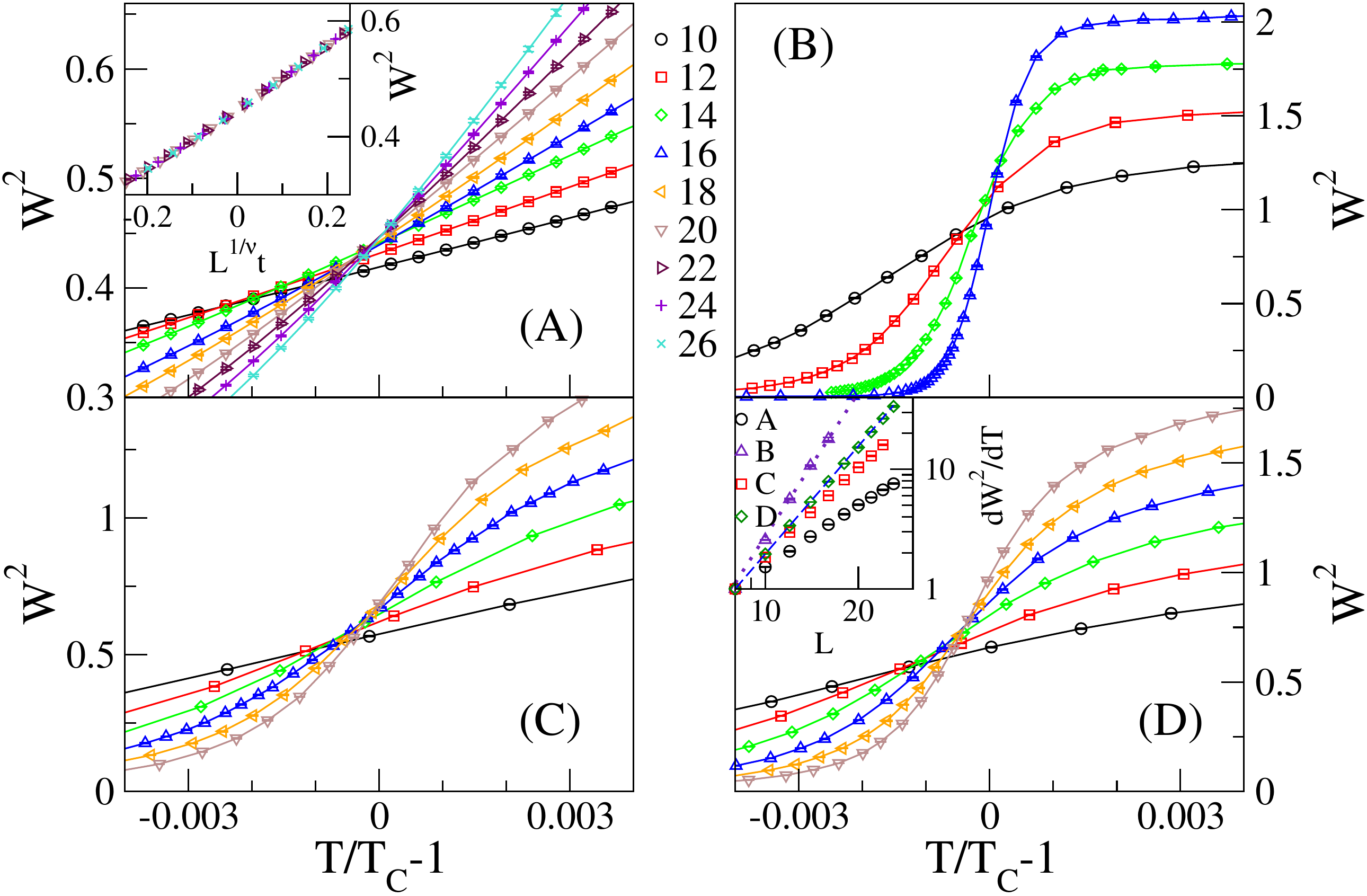}
\caption{(Color online) MC data for vortex loop winding number
  fluctuations.  (A) SR data curves intersect at the critical
  temperature $\tc\approx 1.386$, with scaling corrections visible for
  the smallest system sizes.  Inset: finite size scaling data collapse
  for $L=20,22,24,26$ with $\nu=0.671$.  (B) NM model. The onset is
  steeper than for the SR model, which indicates a first-order
  transition.  (C) SR10 model with $\cdd=-10$.  (D) SR1024999 model
  with $\cdd=-10.24999$. Inset: maximum of the winding number
  fluctuation derivative $dW^2/dT$ for all data sets.  All curves have
  been normalized by the value obtained for $L=8$.  The dashed blue
  line corresponds to a power law $\sim L^3$ and the dotted line to
  $\sim L^4$.}
\label{fig1}
\end{center}
\end{figure}

We start within the strong type-2 regime with model parameters given
by the SR set with a repulsive short-range interaction.
Figure~\ref{fig1} (A) shows MC results for the winding number
fluctuations.  At a second-order phase transition data curves of the
winding number fluctuations vs temperature for different system sizes
$L$ must intersect at the transition temperature according to
Eq.~\eqref{Eq:Wfluctuation}.  Corrections to scaling produce
deviations from the intersection point visible in the figure for the
smallest system sizes, but the biggest sizes intersect within error
bars at a single temperature that estimates $T_c$.  The inset in
Fig.~\ref{fig1} (A) shows a finite-size scaling collapse of MC data
for the four largest system sizes $L=20,22,24,26$ onto a single curve
representing the scaling function $\tilde{W}^2$ in
Eq.~\eqref{Eq:Wfluctuation}.  In the scaling collapse the value
$\nu=0.671$ of the 3DXY model was used.  This is consistent with a
second-order phase transition in the inverted-3DXY universality class
as expected for short range repulsive interactions.

To investigate the effect of a nonmonotonic vortex interaction, MC
data for $W^2$ for the NM parameter set is shown in Fig.~\ref{fig1}
(B).  The data deviates clearly from 3DXY scaling since the slope
$dW^2/dT$ at the transition is much steeper than the 3DXY relation
$L^{1/0.671}$ found in (A).  This demonstrates that the transition of
the NM model is not of the 3DXY type, and it will become clear below
that it is instead first order.  Panels (C) and (D) show data for the
repulsive SR10 and SR1024999 models, respectively.  The inset in (D)
plots the maximum of $dW^2/dT$.  Data curves for the SR and SR10
models both show deviations from a pure power law form for the system
sizes studied here.  The SR data indicates approach to the 3DXY result
$L^{1/0.671}$ for large $L$ but corrections to power-law scaling are
visible also for the largest lattice sizes.  The SR10 model data show
a possible slow crossover towards 3DXY scaling, but the sizes are too
small to decide.  The SR1024999 model is consistent with the size
dependence $dW^2/dT \sim L^3$, showing no tendency for a crossover to
3DXY scaling for the range system sizes examined here.  For the NM
model the data scales approximately as $\sim L^{(d+1)}=L^{4}$.  This
suggests that the assumption of a universal scaling distribution for
the winding number fluctuations $\sim L^{0}$ at the transition does
not hold and indicates that the NM model has a first-order transition.

\begin{figure}[tbp]
\begin{center}
\includegraphics[width=\columnwidth]{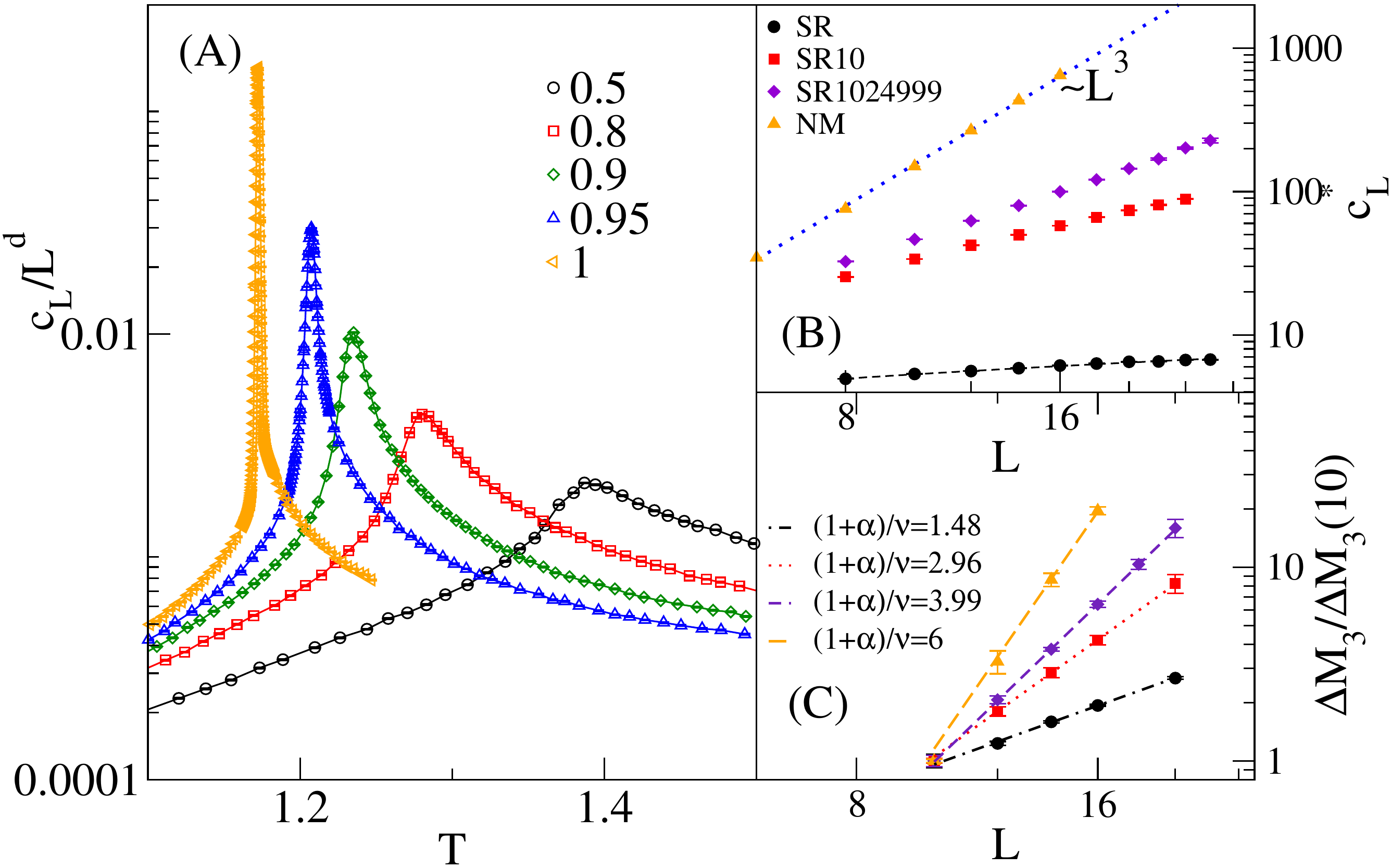}
\caption{(Color online) MC data for the rescaled heat capacity and
  $M_{3}$.  (A) $c_L/L^d$ vs $T$ for $L=14$ for a sequence of models
  ranging from the SR model with $\ldd=0.5$ to the NM model with
  $\ldd=1$.  (B) Scaling of the maxima of $c_L$ vs $L$ for the
  different parameter sets. The dashed black line is a fit to the form
  $aL^{\omega}+b$ with $\omega=-0.02$ and the dotted blue line is a
  pure power law with $\omega=d=3$. (C) Scaling of $\Delta M_{3}$
  vs $L$ normalized by the value at $L=10$.  The NM and SR curves
  scale with the exponents expected for a first-order and
  inverted-3DXY transition, respectively.}
\label{fig2}
\end{center}
\end{figure}

\begin{figure}[tbp]
\begin{center}
\includegraphics[width=\columnwidth]{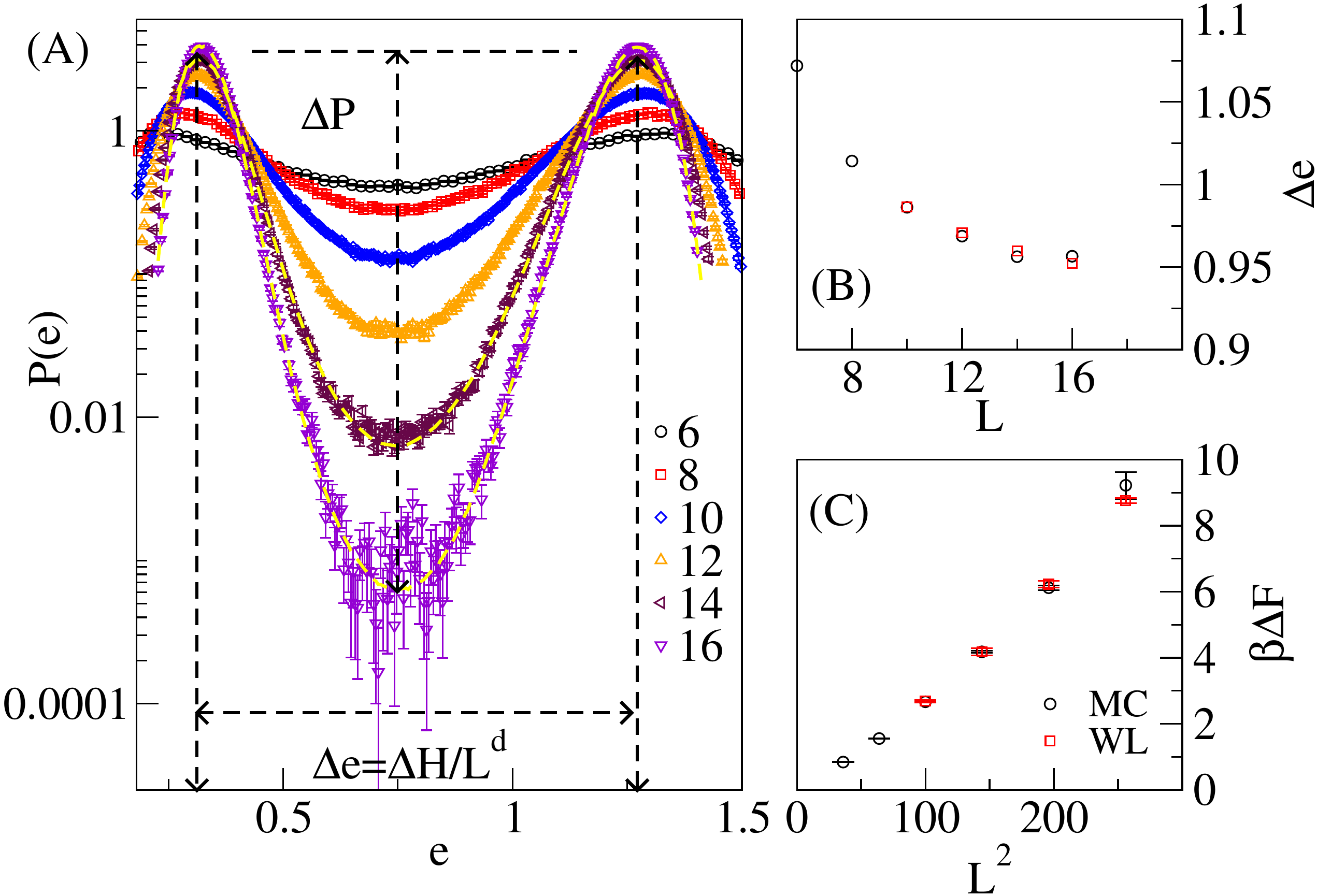}
\caption{(Color online) (A) MC results for the energy histogram at the
  transition for the NM model.  The probability density $P(e)$
  exhibits a characteristic double-peak structure with a monotonically
  growing barrier $\Delta P$ upon increasing the system size.  The
  dashed curves shown for $L=14,16$ are results of Wang-Landau
  calculations.  (B) The normalized latent heat $\Delta e$ obtained
  from the maxima in (A) approaches a nonzero value for $L\ra\infty$.
  (C) The free-energy barrier determined in (A) grows with increasing
  system size which indicates a strong first-order transition.  }
\label{fig3}
\end{center}
\end{figure}

Results for the heat capacity and energy histograms are shown in
Figs.~\ref{fig2} and \ref{fig3}.  Figure~\ref{fig2} (A) shows the
evolution of MC data for the heat capacity $c_L$ for a sequence of
parameter values in $\ldd=0.5-1$ interpolating from the SR to NM case
for system size $L=14$.  The SR data curve is smooth, while increasing
$\ldd$ increases the peak height and decreases the width. The NM curve
peaks sharply at the transition in agreement with a delta-function
peak in the heat capacity at a first-order transition. The insets 
Figs.\ \ref{fig2} (B) and \ref{fig2}(C) show the maximum value $c_L^*$ and the
difference $\Delta M_{3}$ vs $L$.  For the SR model good agreement is
found with $c_L^*\sim L^{2/\nu-3}$ and $\Delta M_{3}\sim
L^{(3-d\nu)/\nu}$ with $\nu=0.671$ as expected for the inverted-3DXY
scenario.  In the NM case the heat capacity maxima scale as $c_L^*\sim
L^d$ and $\Delta M_{3}\sim L^{2d}$ indicating a strong first-order
transition.  Neither of the SR10 and SR1024999 sets show any tendency
towards the same behavior, which suggests either a slow approach to
second order or to weak first-order transitions.  In both SR10 and
SR1024999 cases much bigger system sizes are required for definite
conclusions.

Figure~\ref{fig3} shows energy histograms at the transition
temperature for the NM model and reveals a double-peak structure.  The
results from the Worm and Wang-Landau methods are similar, but the
latter gives smoother data curves in the region between the peaks.
Figure~\ref{fig3} (B) indicates that the latent heat saturates at a
finite value for large system sizes, and (C) shows that the
free-energy barrier grows with increasing system size.  This indicates
that the NM model has a strong first-order transition which is our
main result.

The transition in the models SR10 and SR1024999 is more difficult to
categorize.  Energy histograms for the SR10 model did not produce any
double peaks for the system sizes we explored.  The heat capacity peak
in the inset in Fig.~\ref{fig2} increases significantly slower than a
$\sim L^3$ law expected for a first-order transition, and may possibly
approach 3DXY scaling for large systems for the SR10 model.  Both
these results favor a second-order transition.

\begin{figure}[tbp]
\begin{center}
\includegraphics[width=\columnwidth]{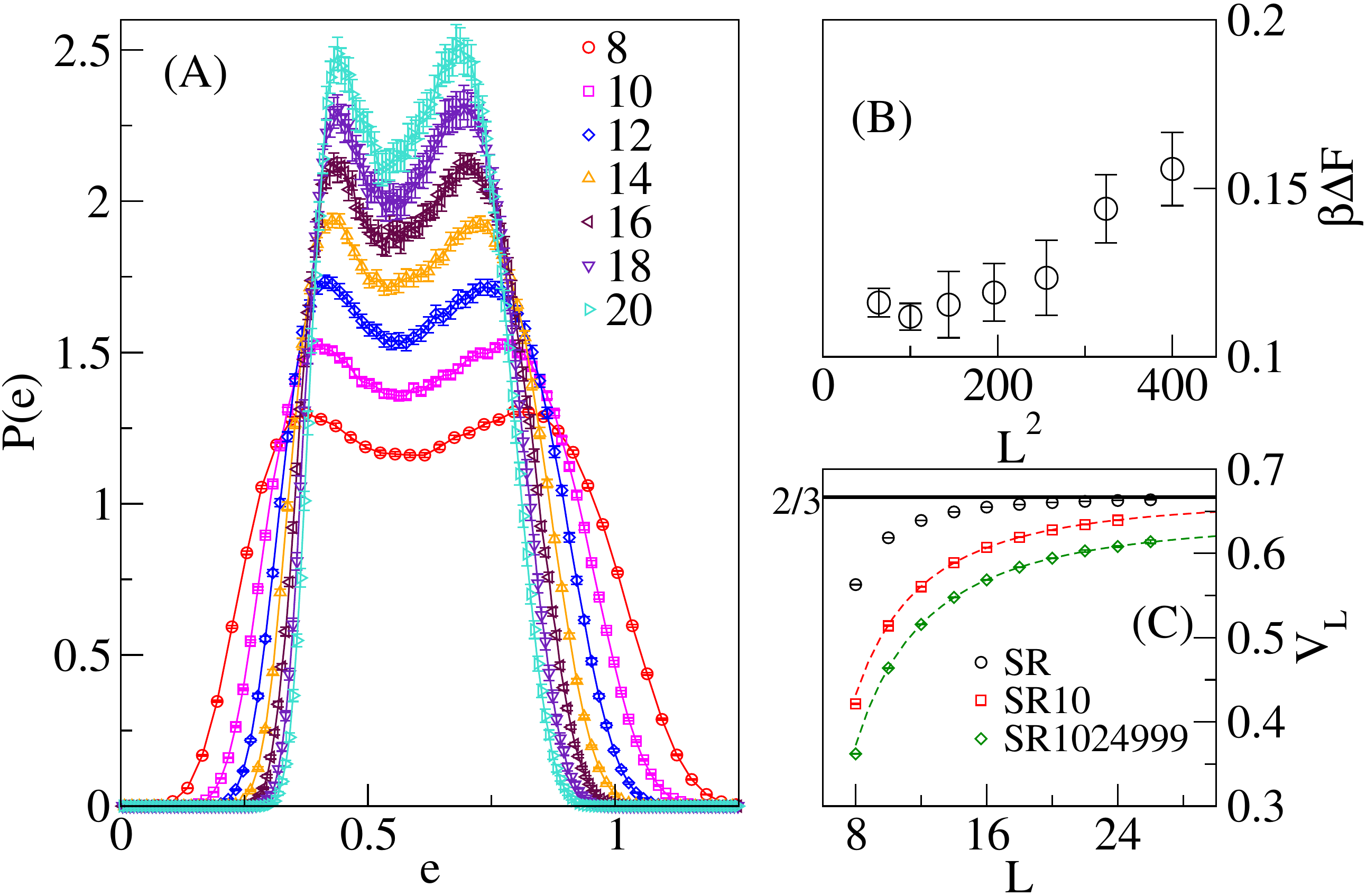}
\caption{(Color online) (A) Energy histogram of the SR1024999 model at
  the phase transition. The tendency towards double-peak formation is
  much weaker than for the NM case plotted in Fig.~\ref{fig3}.  (B)
  The free-energy barrier $\Delta F$ increases slowly with system
  size. (C) Minimum value of the energy cumulant $V_L$ vs system size
  $L$.  The black line is the asymptotic limit 2/3 expected for second
  order transitions.  Dashed curves are fits to the form
  $V_L=V_\infty+a/L^b$.  }
\label{fig4}
\end{center}
\end{figure}

Figure~\ref{fig4} shows energy histogram data for the SR1024999 model.
Figure~\ref{fig4} (A) shows a double-peak structure in the histogram. Figure~\ref{fig4} (B) shows a free
energy barrier growing slowly with increasing system size which is
expected at a first-order transition.  However, the heat capacity
maximum plotted in Fig.~\ref{fig2} (B) grows slower than a $L^d$
power law corresponding to a first-order transition, indicating that
the width of the energy histogram $\Delta e \to 0$ for $L\to\infty$.
While this suggests a continuous transition, two further
observations can be made.  The scaling deviation from first-order
behavior could in principle be attributed to finite size corrections
as follows.  The width of the energy histogram in Fig.~\ref{fig4} (A)
is related to the heat capacity data in Fig.~\ref{fig2} by $\Delta e_L
\sim (L^{-d} c_L)^{1/2}$.  A finite size scaling ansatz $\Delta
e_L=\Delta e_\infty+A/L+B/L^2$ with fit parameters $\Delta e_\infty,
A, B$, gives a good fit to the data and extrapolates to a finite peak
width $\Delta e_\infty \approx 0.18$ in the large system limit, which
is consistent with a first-order transition showing substantial finite
size corrections.  Alternatively, a power law of the form $\Delta e_L
\sim 1/L^p$ also fits the data and gives $p \approx 0.7$, which
extrapolates to a single delta peak histogram for $L \to \infty$.
This corresponds to a heat capacity maximum that varies with system
size as $c_L^* \sim L^{1.6}$.  However, as also seen in
Figs.~\ref{fig2} (B) and (C), this is far from the finite-size scaling
behavior expected at a 3DXY transition given by $c_L^* \approx
aL^{-0.02}+b$.

Figure~\ref{fig4} (C) shows results for the minimum of the energy
cumulant $V_L$.  All data curves can be fitted to the form $V_L =
V_\infty + aL^{-b}$.  Data for the SR model quickly converges towards
the expected value $V_\infty=2/3$ for a second-order transition.  A
fit of the data for the SR10 model gives $V_\infty=0.66\pm0.01$ which
is consistent with a second-order transition.  The corresponding fit
for the SR1024999 model gives a slightly lower asymptotic limit
$V_\infty=0.64\pm0.01$, but with a correction exponent $b\approx 1.9$
rather than $b=d=3$ expected for a strong first-order transition.  The
NM model gives a negative minimum value of $V_{L}$ for all $L$
consistent with a strong first-order transition (data not shown).

\begin{figure}[tbp]   
\begin{center}
\includegraphics[width=\columnwidth]{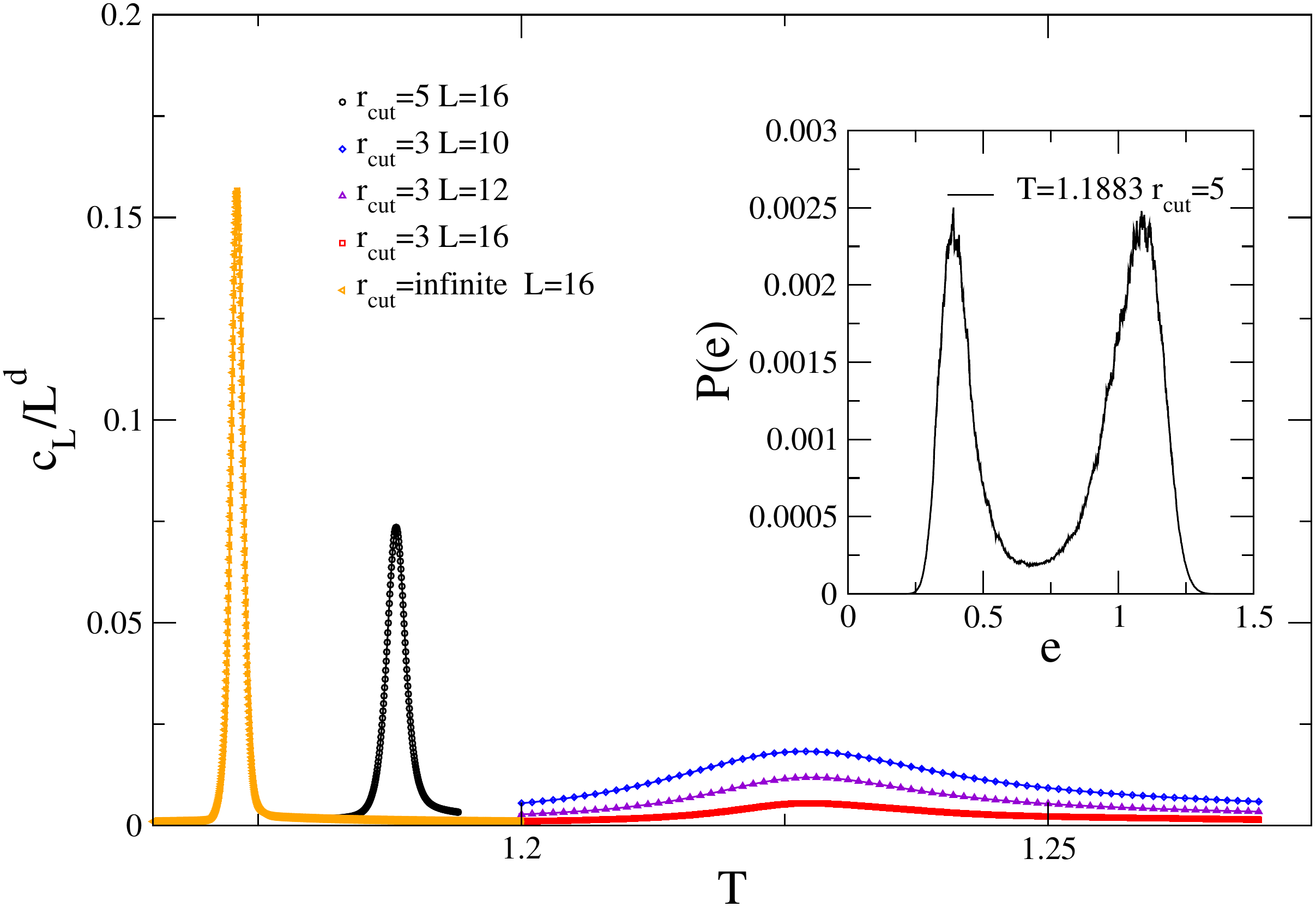}
\caption{(Color online) Rescaled heat capacity for different cutoff
  lengths of the NM model interaction.  The NM model without
  truncation (orange curve) has a first-order transition with a sharp
  peak in the heat capacity.  The truncated model with $r_c=3$ (blue,
  purple, red curves) has a monotonic interaction, and gives a
  rescaled heat capacity peak that decreases with system size
  indicating a second-order transition.  For the nonmonotonic model
  with $r_c=5$ (black curve) first-order signatures reappear, i.e., a
  sharp heat capacity peak, and a double-peaked energy histogram at
  the transition (inset).  }
\label{fig5}
\end{center}
\end{figure}

To further assess the importance of nonmonotonicity in the vortex
interaction we implemented the following modification of the NM model.
The attractive potential $U_{ij}$ in Eq.~\eqref{Eq:ModelHam} of the NM
model was set to zero beyond a cutoff radius $r_c$ without altering
the interaction for $r<r_c$.  Taking the cutoff radius to $r_{c}=3$
gives a totally repulsive interaction in all directions.  The effect
of truncating the interaction at different distances is demonstrated
in Fig.~\ref{fig5}, which indicates that for $r_c=3$ the first-order
behavior vanishes and the transition becomes second order, presumably
turning into a 3DXY transition as for the SR case.  The energy
histograms in this case have no double-peak structure.  If the cutoff
radius is chosen to $r_c=5$ the interaction becomes nonmonotonic in
all directions.  Then the first-order signature reappears as shown by
the black curve and the double-peaked energy histogram in the inset in
Fig.~\ref{fig5}.  This indicates that the first-order mechanism found
in the NM model is affected by the nonmonotonicity of the vortex
interaction.

Taken together the results in Figs.\ \ref{fig1}, \ref{fig2} and \ref{fig5} 
suggest $d\nu>1$ for all models with a screened monotonic interaction.
This  implies a collapse of the histogram in Fig.\ \ref{fig4} to a single peak in 
the thermodynamic limit, and thus second-order transitions for all SR models.
However, given the results
presented in Fig.\ \ref{fig4} and the fact that the exponents for the
SR10 and SR1024999 data in Figs.\ \ref{fig1} and \ref{fig2} clearly
deviate from the 3DXY values, weak first-order transitions in the
thermodynamic limit cannot be completely ruled out.  The model then
would have weak first-order transitions also for parameters where the
interaction between vortices is fully repulsive.  Again, in this near
degenerate regime the model does not correspond to ordinary type-2
superconductors.

\section{Discussion}

We present simulation results suggesting that in type-1.5
superconductors there is a new mechanism that drives the
superconducting transition to become first order.  This however does
not imply that the superconducting phase transition in type-1.5
material is generically first order.  We described the fluctuations by
a generalized effective link-current model. To answer the question in
the general case requires much more computationally demanding
large-scale simulations of full two-band Ginzburg-Landau models. It is
conceivable that fluctuation-induced enhancement of the repulsion for
some of the parameters of the model eliminates the bare attractive
interaction between vortices, which may make the phase transition
continuous for certain parameter ranges in the type-1.5 regime.  Among
various scenarios for realization of type-1.5 superconductivity, a
special reservation should be made for simple $U(1)$ multiband
materials. In that particular case, at the mean-field level, the
intervortex interaction depends on temperature and a superconductor
becomes either type-1 or type-2 in the limit $T\to T_c$ by standard
mean-field symmetry-based arguments \cite{silaevmicro}. This is
consistent with experiments that study the temperature dependence of
the vortex attraction \cite{ray}.  Therefore for this particular kind
of type-1.5 materials for the phase transition to be first order, the
fluctuations should be strong enough so that the phase transition
takes place substantially below the mean-field estimate of $T_c$. This
means that the effects suggested here are probably more likely to be
observed in multiband type-1.5 superconductors with a relatively high
$T_c$.

\begin{acknowledgments}

  We thank Jack Lidmar for valuable discussions.  E.B.\ was supported by
  the Knut and Alice Wallenberg Foundation through a Royal Swedish
  Academy of Sciences Fellowship, by the Swedish Research Council
  Grants No.\ 642-2013-7837 and No.\ 325-2009-7664. Part of the work was done at
  University of Massachusetts Amherst and supported by the National
  Science Foundation under the CAREER Grant No.\ DMR-0955902.  M.W.\ was
  supported by the Swedish Research Council Grant No.\ 621-2012-3984.
  Computations were performed on resources provided by the Swedish
  National Infrastructure for Computing (SNIC) at PDC and at the
  National Supercomputer Center in Link\"oping, Sweden.

\end{acknowledgments}

\input{nonmonotonic.bbl}
\end{document}

%% file: nonmonotonic.bbl
%